\documentclass[11pt]{article}

\usepackage{epsfig}
\usepackage{latexsym}
\usepackage{enumerate}
\usepackage{url}
\usepackage{float}
\usepackage[hypertexnames=false,hyperfootnotes=false]{hyperref}
\usepackage{texnansi}
\usepackage{color}
\usepackage{tikz}
\usepackage[margin=10pt,font=small,labelfont=bf]{caption}
\usepackage[subrefformat=parens,labelformat=parens]{subcaption}
\usepackage{afterpage}
\usepackage{enumitem}
\usepackage[boxed]{algorithm}
\usepackage{algpseudocode}
\usepackage[normalem]{ulem}
\usepackage{lmodern}
\usepackage{booktabs}
\usepackage{sectsty}
\usepackage{ifthen}
\usepackage{amsmath}
\usepackage{amsthm}
\usepackage{amssymb}
\usepackage{amsfonts}
\usepackage{thmtools}
\usepackage{thm-restate}
\usepackage{mathtools}
\usepackage{xspace}
\usepackage{titling}
\usepackage{natbib}
\usepackage{xfrac}
\usepackage{multirow}
\usepackage{bigdelim}
\usepackage{bm}
\usepackage{cleveref}
\usepackage[textsize=tiny]{todonotes}
\declaretheoremstyle[headfont=\sffamily\bfseries,bodyfont=\itshape]{thm-sf}

\crefname{assumption}{assumption}{assumptions}

\renewcommand{\proofname}{{\normalfont\sffamily\bfseries Proof}}

\renewcommand{\thmcontinues}[1]{\hyperref[#1]{continued}}
\newcommand{\paraheader}[1]{\smallskip\noindent{\sffamily\bfseries #1}}
\usetikzlibrary{arrows,patterns,plotmarks,pgfplots.groupplots}
\tikzstyle{every picture} += [>=stealth]
\tikzset{axis/.style={semithick, line join=miter}}
\allsectionsfont{\sffamily}
\makeatletter
\def\@seccntformat#1{\csname the#1\endcsname.\quad}
\makeatother
\renewcommand{\figurename}{\normalfont\sffamily\bfseries Figure}
\renewcommand{\tablename}{\normalfont\sffamily\bfseries Table}
\renewcommand{\abstractname}{\normalfont\sffamily\bfseries Abstract}
\floatname{algorithm}{\normalfont\sffamily\bfseries Algorithm}
\floatstyle{ruled}
\newcommand{\emailhref}[1]{\href{mailto:#1}{\tt #1}} 
\provideboolean{fastcompile}
\newcommand{\hidefastcompile}[1]{\ifthenelse{\boolean{fastcompile}}{}{#1}}
\usepackage{pgfplots}
\usepackage{pgfplotstable}
\usetikzlibrary{calc}
\usepackage{mathtools}
\definecolor{orange}{rgb}{0.85,0.33,0.13} 
\definecolor{green}{rgb}{0.13,0.85,0.33}
\definecolor{purple}{rgb}{0.33,0.13,0.85}
\definecolor{lime}{rgb}{0.65,0.85,0.13}
\definecolor{blue}{rgb}{0.13,0.65,0.85}
\pgfplotscreateplotcyclelist{tricolor}{%
  orange,every mark/.append style={fill=orange!80!black},mark=*\\%
  green,every mark/.append style={fill=green!80!black},mark=square*\\%
  purple,every mark/.append style={fill=purple!80!black},mark=otimes*\\%
  black,mark=star\\%
  orange,every mark/.append style={fill=orange!80!black},mark=diamond*\\%
  green,densely dashed,every mark/.append style={solid,fill=green!80!black},mark=*\\%
  purple,densely dashed,every mark/.append style={solid,fill=purple!80!black},mark=square*\\%
  black,densely dashed,every mark/.append style={solid,fill=gray},mark=otimes*\\%
  orange,densely dashed,mark=star,every mark/.append style=solid\\%
  green,densely dashed,every mark/.append style={solid,fill=green!80!black},mark=diamond*\\%
}
\pgfplotsset{colormap={tricolormap}{color=(orange) color=(green) color=(purple)},
  colormap={quadcolormap}{color=(orange) color=(lime) color=(blue) color=(purple)}}
\pgfplotstableset{%
  font=\small,
  every head row/.style={before row=\toprule[1pt], after row=\midrule},
  every last row/.style={after row=\bottomrule[1pt]}}
\pgfplotsset{compat=1.15}

\usepackage{setspace}
\usepackage[margin=1in]{geometry}

\newcommand{\LVR}{\ensuremath{\mathsf{LVR}}\xspace}

\title{\bf\sffamily Layer 2 be or Layer not 2 be: Scaling on Uniswap v3}
\author{
Austin Adams\thanks{
 We thank Maude Wilson, Gordon Liao, Ciamac Moallemi, Ed Felten, Brad Bachu, Xin Wan, and Bridgett Frey for their comments.} \\ Uniswap Labs \\ \emailhref{austin@uniswap.org} 
}

\begin{document}
\maketitle
\singlespacing

\begin{abstract}
This paper studies the market structure impact of cheaper and faster chains on the Uniswap v3 Protocol. The Uniswap Protocol is the largest decentralized application on Ethereum by both gas and blockspace used, and user behaviors of the protocol are very sensitive to fluctuations in gas prices and market structure due to the economic factors of the Protocol. We focus on the chains where Uniswap v3 has the most activity, giving us the best comparison to Ethereum mainnet. Because of cheaper gas and lower block times, we find evidence that the majority of swaps get better gas-adjusted execution on these chains, liquidity providers are more capital efficient, and liquidity providers have increased fee returns from more arbitrage. We also present evidence that two second block times may be too long for optimal liquidity provider returns, compared to first come, first served. We argue that many of the current drawbacks with AMMs may be due to chain dynamics and are vastly improved with cheaper and faster transactions.
\end{abstract}

\onehalfspacing

\section{Introduction}

Automated market makers (AMMs) - like the Uniswap Protocol - have become the dominant market structure for onchain markets, trading over \$3 trillion dollars to date.\footnote{\href{https://defillama.com/dexs}{Link to source}} The Ethereum Protocol, currently the largest chain by onchain volume, has a limited availability for computation - or block space - and as user demand increases, the cost has risen due to limited supply. This is by design - block space needs to be constrained due to security and \href{https://x.com/notnotstorm/status/1764700326282879115?s=20}{state growth}.  But as markets mature and the average transaction \href{https://twitter.com/bantg/status/1692467461952582087?s=20}{becomes more complex} (increasing aggregate transaction costs), \href{https://etherscan.io/chart/gaslimit}{Ethereum's} base-layer throughput cannot feasibly keep up. As a result, there are comparatively fewer transactions on mainnet today than a few years ago, and those transactions are more expensive than they used to be.

To scale Ethereum, the \href{https://twitter.com/VitalikButerin/status/1588669782471368704}{ecosystem} has chosen a modular scaling approach through utilizing layer 2 protocols (L2s). L2s utilize the trustless data availability of Ethereum while outsourcing computational validation to various schemes, such as zero-knowledge or optimistic fraud proofs. Network costs on L2s are much lower compared to Ethereum mainnet, and that cost will reduce even further with the forthcoming implementation in \href{https://github.com/ethereum/consensus-specs/blob/dev/specs/deneb/beacon-chain.md}{Dencun} of \href{https://eips.ethereum.org/EIPS/eip-4844}{proto dank-sharing}.\footnote{Dencun refers to the joint consensus hardfork \href{https://github.com/ethereum/consensus-specs/blob/dev/specs/deneb/beacon-chain.md}{Deneb} and execution hardfork \href{https://github.com/ethereum/execution-specs/blob/master/network-upgrades/mainnet-upgrades/cancun.md}{Cancun}, which goes by the portmanteau, Dencun} At the current time, \href{https://twitter.com/VitalikButerin/status/1764139239233749047}{prediction markets} price a 60x decrease in the already cheap data availability costs of L2s, but the benefit will likely be lower due to induced demand.

In this paper, we will study the impact of cheaper and faster transaction costs on the Uniswap Protocol. To do this, we will contrast market structure of Uniswap v3 on largest L2s to Ethereum mainnet. We will focus on L2s, because the largest deployments of Uniswap v3 are on these chains and Ethereum mainnet. However, we should note that sidechains or alternative L1s - like Solana or Monad - have some of the same benefits.

Lower costs will have significant benefits for Uniswap Protocol users. According to \cite{adams2023costs}, 22\% of the total cost of the average Ethereum transaction on the Uniswap Interface is Ethereum gas costs. On transactions less than \$1k, gas costs can make up as much as 98\% of the total costs of trading. Lowering this cost allows users to pay less for the same service, potentially making trading costs comparable to centralized exchanges. Additionally, cheaper arbitrage transactions, and the related increase in fees paid by arbitrageurs, make LPs more profitable. It is also possible that users - both swappers and LPs - who are currently priced out of Ethereum mainnet would take part in the ecosystem if it were cheaper. 

Today, AMMs on Ethereum currently have better execution for large trades due to higher market depth (because of higher TVL) and amortized gas costs over the larger input amount. However, most users (and future potential users) likely do not trade the size where it is better to trade on Ethereum. By decreasing costs, we can show the potential of scaling beyond Ethereum, and the practical benefits of AMMs to new users by allowing cheaper, more efficient, and easier transactions.

\medskip
\paraheader{Contributions.} 
In this paper, we contribute to the blossoming field of Uniswap v3 empirical research by focusing on the impact cheaper and faster chains have on the largest application by blockspace on Ethereum - the Uniswap Protocol. We show that the around 90\% of Uniswap Interface swappers would have better execution on the studied L2s. LPs on the Protocol also benefit from 50\% average higher capital efficiency, and 20\% more average arbitrage fee returns. We also talk about how these differences could alleviate many of the concerns for wide-spread adoption of AMMs outside of the crypto ecosystem.

There is currently a lack of research on the impact of faster chains on Uniswap v3, because the required data is cumbersome to pull, maintain, and utilize. To support more research in this area, we developed and open-sourced a Python package that recreates Uniswap v3 functionality natively in Python - called v3-polars - along with this paper. It can be \href{https://github.com/Uniswap/v3-polars}{found here}. All empirical work in this paper is either created using this package and can be found in \href{https://github.com/aadams/layer2-be}{this repository}, or the link is provided. Therefore, the entire paper should be replicable. 

Overall, we present the first overview of the entire Uniswap v3 ecosystem on Ethereum mainnet and L2s and compare them. We discuss the current benefits - and the downsides - of the current L2 ecosystem for the Uniswap Protocol. As Ethereum continues to move towards modularity, we hope this paper provides insight into what the future of onchain decentralized markets may look like, and how these changes could solve some of the challenges with widespread AMM adoption today.

\medskip
\paraheader{Related Literature.} 
Literature on automated market makers has grown rapidly in the past few years. Early roots can be traced back to \cite{hanson2007logarithmic} and \cite{othman2013practical}. Some recent literature about automated market makers include \citet{capponi2021adoption}, \citet{lehar2021decentralized}, and \cite{lvr2022}. Uniswap v3 specifically has been described in \cite{Adams21} and empirically studied by \cite{adams2023costs}.
	
While there are relatively few papers on the empirics of Uniswap v3 outside of Ethereum mainnet, most likely due to the difficulty of generating the necessary data, there are a few notable contributions. \cite{chemaya2022cost} presents findings showing that CPMMs on L2s give users better prices compared to Ethereum, but that users do not optimally switch based on the best total cost - showing a perceived security metric for the chain. \cite{caparros2023blockchain} focuses on slippage and concentration between Polygon and Ethereum, showing similar benefits from downstream gas cost reductions. \cite{mitchell2022layer2} shows that the average full-range liquidity position does not have increased total fee returns on L2s, which will not contradict our results - as we focus only on arbitrage fee returns.

A technical discussion of L2s is out of the scope of this paper. For that, see \cite{thibault2022blockchain}.


\section{Overview of Current Uniswap v3 Deployments}

In this paper, we will focus on the Ethereum mainnet, the two rollups with significant Uniswap v3 deployments (Arbitrum and Optimism), and the sidechain, Polygon. Some argue that Polygon is not technically an L2 due to a distinct validator set, but we will refer to it as an L2 for brevity. While Uniswap v3 is deployed on other chains, these chains have the most volume and TVL of the Uniswap v3 deployments.\footnote{For an exhaustive and updated list of canonical deployments, see \href{ https://app.ens.domains/v3deployments.uniswap.eth}{here}} The majority of volume is still found on mainnet, and during times of volatility volume is proportionally higher. However, it is important to keep in mind that although volume may be higher on mainnet, the majority of transactions actually occur on L2s. Ethereum generally only has about 25\% share of transaction count, but around 60\% of volume.

\begin{figure}[H]
\centering
\textbf{Figure 1A: Share of Volume by Blockchain on Uniswap v3
}\par\medskip 
\includegraphics[width=.8\textwidth]{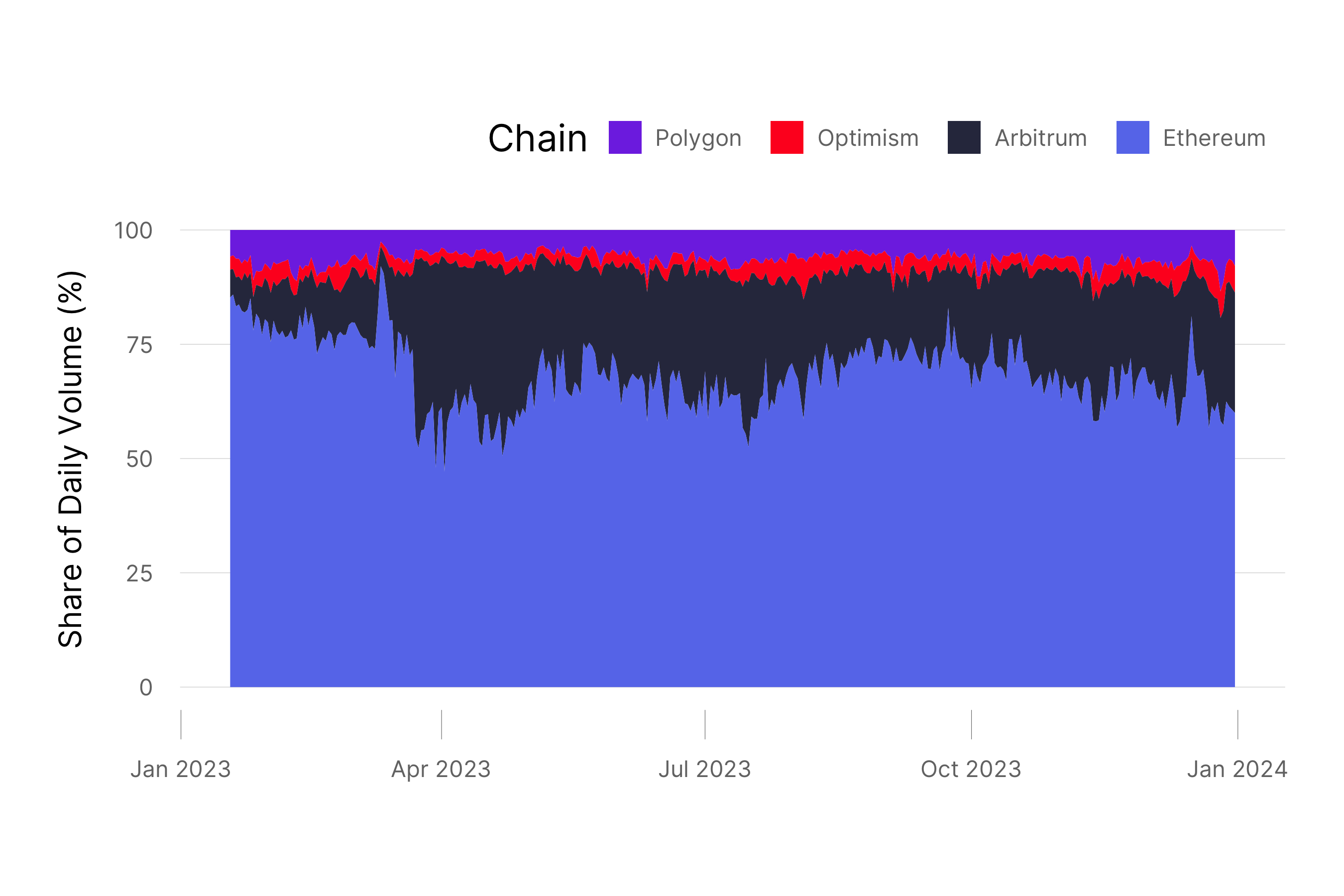}
\caption*{\href{https://dune.com/queries/2318904/3795564}{Source}}
\end{figure}

\begin{figure}[H]
\centering
\textbf{Figure 1B: Share of Transactions by Blockchain on Uniswap v3
}\par\medskip 
\includegraphics[width=.8\textwidth]{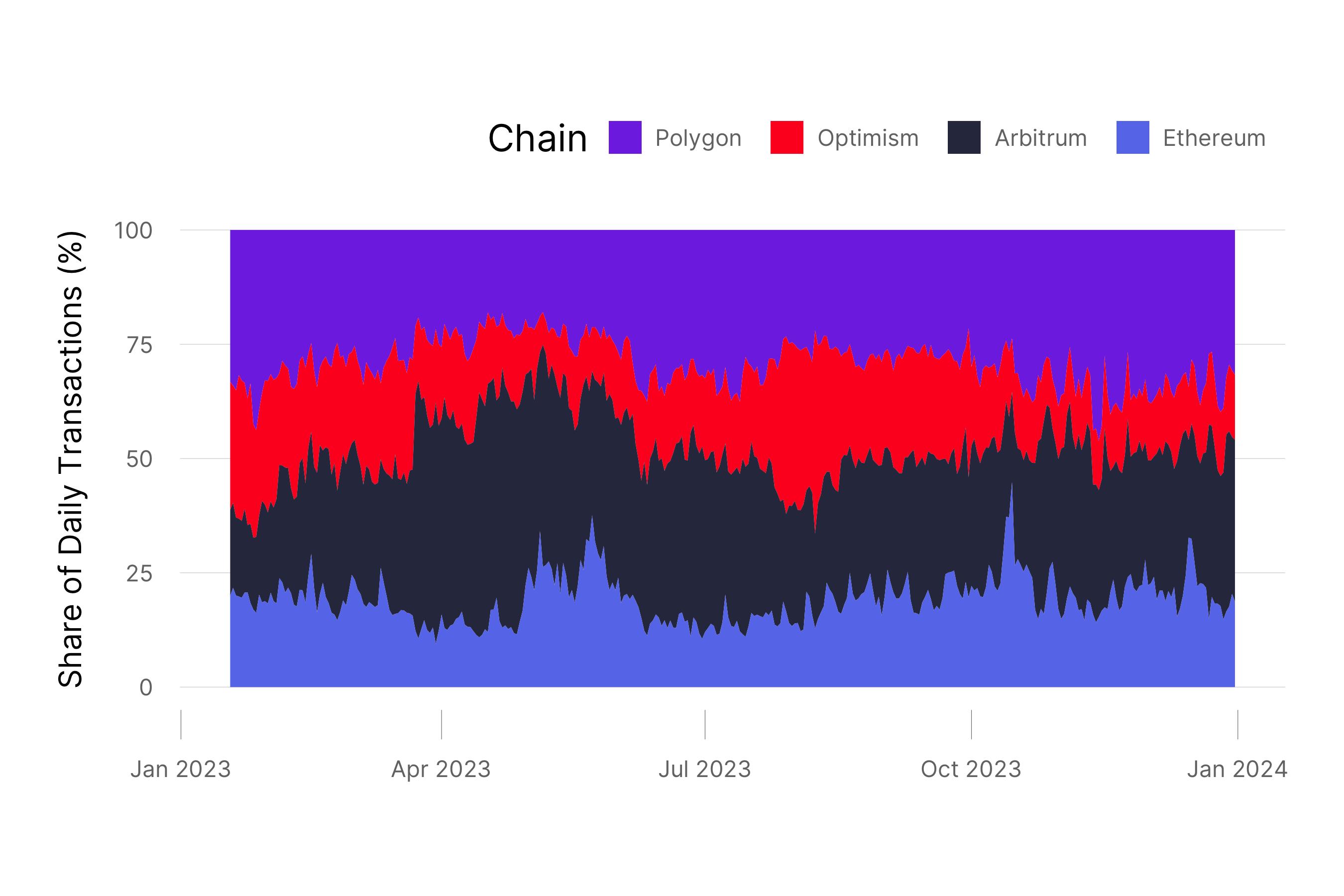}
\caption*{\href{https://dune.com/queries/3435845/5771579}{Source}}
\end{figure}

During times of high volume, volume returns proportionally to Ethereum mainnet, as Ethereum’s share of total volume spikes during high volume days (which can be seen in the spikes of Ethereum market share in Figure 1A). There are few potential reasons for this. First, when there is volatility, the larger value locked on Ethereum mainnet has greater turnover, resulting in a more extreme volume spike. Users (or likely arbitrageurs) may also be more inelastic to gas costs during times of volatility, making them more likely to swap at high gas costs due to the presence of profitable arbitrage opportunities resulting from volatility. This negates one of the major benefits of L2s, leading to increased volume on mainnet.

\begin{figure}[H]
\centering
\textbf{Table 1: Uniswap v3 Protocol Statistics}
\par\medskip 
\includegraphics[width=.8\textwidth]{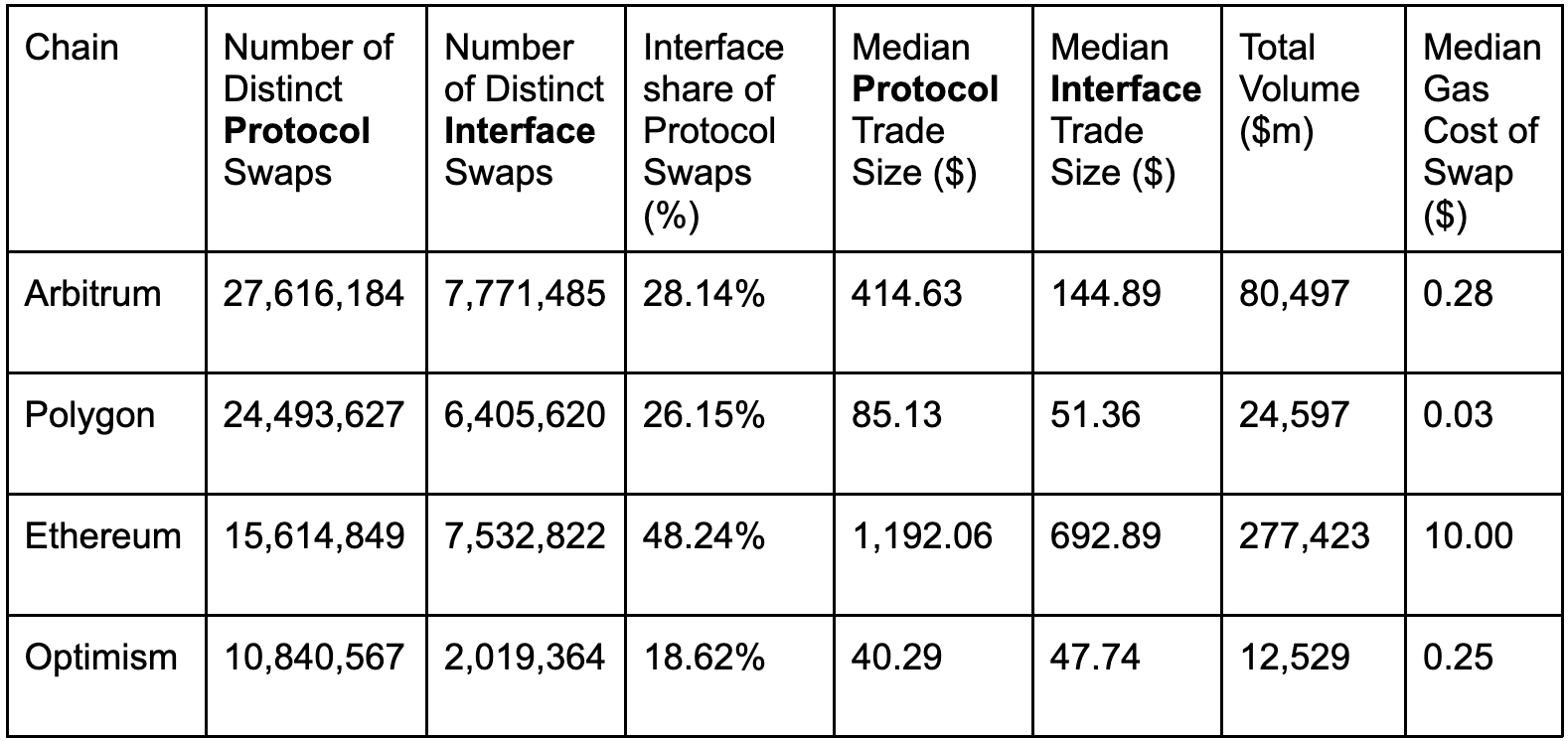}
\caption*{Note: All calculations are year to date (Jan 1st, 2023 - Jan 1st, 2024) \\
Note: Ethereum statistics only include Uniswap v3 (to allow direct comparisons) \\
\href{https://dune.com/queries/3440221}{Source}}
\end{figure}

Any analysis of Protocol volume must look at both the amount of trades being executed and the size of those trades to get the complete picture. While the majority of volume is still found on mainnet, the majority of transactions are actually occurring on L2s. While Arbitrum has almost twice as many swaps as mainnet, the median size of mainnet swaps is almost three times larger than Arbitrum, leading to higher aggregate volume on Ethereum. One explanation for this is the higher liquidity on mainnet, and the arbitrage that occurs is larger as a result. More liquidity leads to larger arbitrage transactions, but not more swapping opportunities, leading to a larger average swap size from arbitrage. The other main reason is that fixed gas costs for swaps on mainnet are larger, and people are less likely to execute small transactions with high gas costs. 

Historically, Ethereum has dominated the number of users and volume trading on it likely due to first mover advantage and censorship resistance properties. However, retail users likely benefit the most from L2s because of lower gas costs and better average execution. The main market structure benefit of Ethereum is the large capital base in pools. We expect to see price-sensitive users begin to swap on cheaper chains as adoption continues, especially if their drawbacks are improved.


\section{Benefits for Uniswap v3}

There are two main categories of swappers on blockchain-based markets: arbitrageurs and retail flow. Aggregate statistics combine these two users, although they have different reasons for trading. Arbitrageurs will execute orders if they are profitable, and we will revisit their impact on market structure later. Retail flow is just as important (providing the bulk of positive returns for LPs in the \LVR framework established in \cite{lvr2022}) and are a bulk of the users who benefit from cheaper chains.

To focus on retail users, we must first segment out arbitrageurs. To help us segment, we will take advantage of the fact that arbitrageurs do not typically arbitrage using the Uniswap Interface. The Interface contracts were designed to support safe onchain trading, and are typically less gas efficient and less customizable than optimally designed MEV contracts, which are preferred by arbitrageurs. Retail users are less gas sensitive than MEV traders, preferring better trade outcomes on the relatively fewer swaps they do, so they are more likely to use the Interface. Because of this, we can segment out the Interface users as a separate cohort of predominantly retail users.\footnote{Note that while arbitrageurs do not use the Uniswap Interface, not all retail flow uses the Uniswap Interface. This means that we are likely dropping significant retail flow from our sample and should not be taken as exhaustive.} We will utilize this same segmentation feature in reverse in later sections. 

We can see that as a percentage of total Interface transactions, trades on Ethereum are almost double the next highest chain (around 48\% for Ethereum to 28\% on Arbitrum). These retail transactions are smaller than arbitrage transactions, but still larger than retail transactions on other chains. 

Arbitrage transactions are likely lower bounded by gas costs, as arbitrageurs will only execute if they are able to pay off their fixed costs for transacting (gas). However, Uniswap Interface volume has no implicit lower-bound, but even these trades on the Uniswap Interface on Ethereum are bigger. This could indicate that users who utilize lower trade sizes must move off of Ethereum and are effectively priced out. We will focus on costs for users to find the optimal switching point.

There are two main costs to users, gas costs and price impact. From \textbf{Table 1}, we can directly calculate the gas cost to users. The median swap gas cost over the last year on Uniswap v3 on Ethereum mainnet is \$10.00, while the average Arbitrum swap cost is \$0.28 - a staggering 97\% decrease in gas cost. 

The other cost of swapping for users is price impact, which is the movement of the “midmarket price” of the pool from execution of the trade. Uniswap v3 allows users to “concentrate liquidity” between prices, concentrating liquidity compared to a CPMM (like Uniswap v2). This concentration of liquidity has led to market depth dwarfing that of centralized exchange in some cases. For example, Uniswap v3 has two times the market depth of ETH/USD versus Binance - see \cite{liao2022dominance} for more discussion on this.

\medskip
\paraheader{Small Swappers Benefit From Lower Gas.} 

First, we will examine the difference in all-in swap cost of an USDC to ETH trade on Ethereum mainnet vs. Arbitrum. As previously stated, we will break down the cost into two factors: price impact and gas costs. We saw in \textbf{Table 1} that gas costs on the studied L2s are much smaller, however overall liquidity is lower (and thus the price impact of a trade is higher). We utilize similar methodology to \cite{chemaya2022cost} who calculate the optimal time for switching between chains.

\begin{figure}[H]
\centering
\textbf{Figure 2: Breakeven size of ETH/USDC on Ethereum vs. Arbitrum
}\par\medskip 
\includegraphics[width=.8\textwidth]{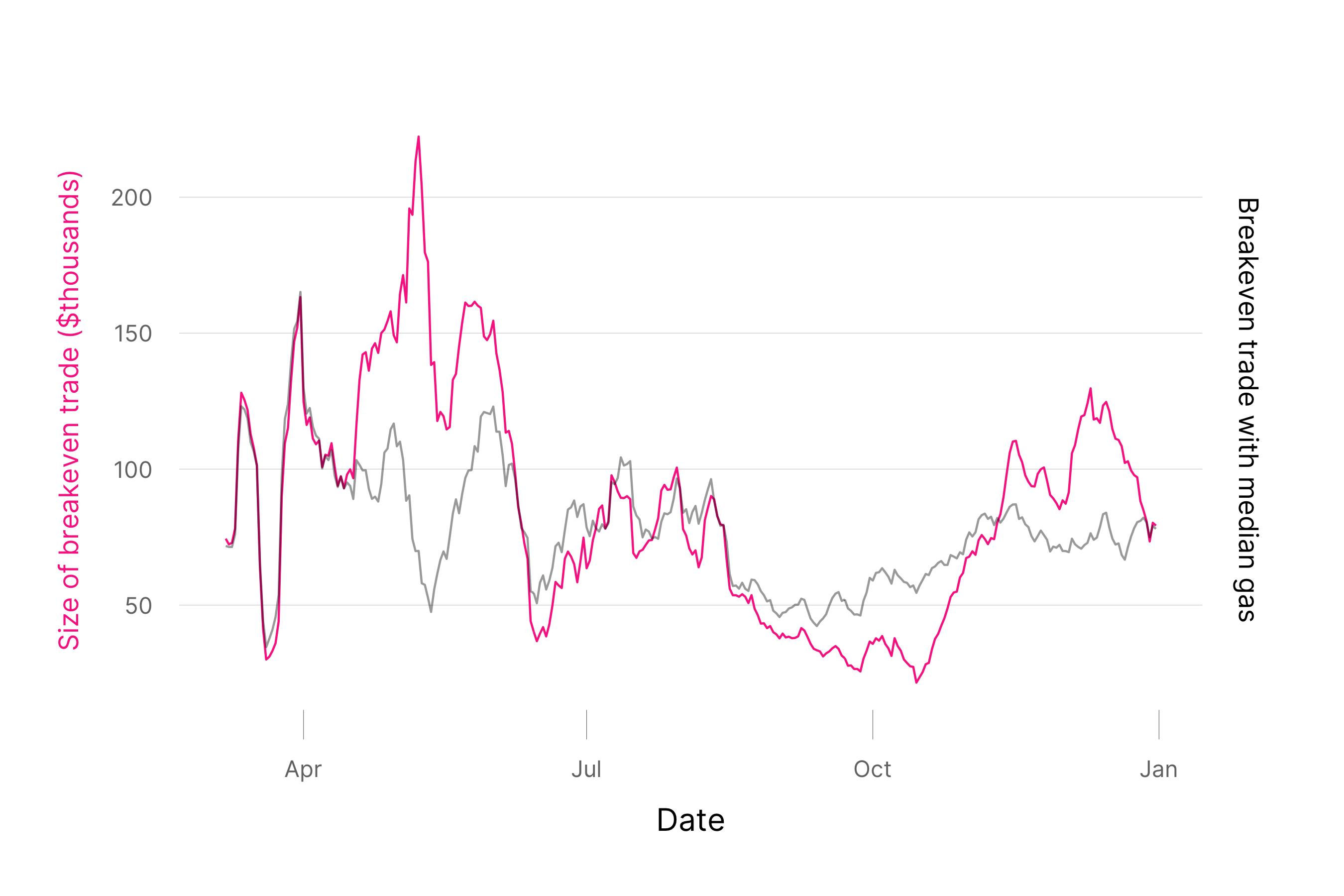}
\caption*{Note: The above calculation does not take into account the increase in gas cost of a higher swap size (due to the cost of shifting ticks). Since these costs are logarithmic to the swap size and a small portion of the overall gas cost, they should not notably impact these values, but slightly higher than calculated. The grey line represents the breakeven cost at each time if gas was always the median value over the sample. We also focus only on Arbitrum, because of non-standard gas calculations with Optimism.}
\end{figure}

In \textbf{Figure 2}, we examine the trade size in dollars of an USDC to ETH swap that is needed to breakeven - that is a swap of USDC to ETH where we would get an equal output of ETH on Arbitrum and Ethereum (including gas). In summary, above the pink line a user receives more ETH on Ethereum and below would receive more ETH on Arbitrum. The grey line calculates the same value, but holds gas fixed to the median value over the time period to show the impact of price impact.

First, let us focus on the pink line which utilizes gas costs as of the time of the swap. Notice that the size of this swap varies drastically over the last year. To pick a number as a baseline, we choose \$25k, as breakeven is strictly over this value. To put this into context, around 10.9\% of swaps that utilize the ETH/USDC pool from the Uniswap Interface are above \$25k during this time period.\footnote{\href{https://dune.com/queries/3263635}{Source}} Only 2.5\% of swaps are greater than \$125k, which is the breakeven value at the end of the sample period. These numbers indicate that the vast majority of users do not trade with the transaction size needed to benefit from the larger capital base on Ethereum, meaning they are likely better of mainly operating on cheaper chains. Note that there are more swaps that go through these pools than just Uniswap Interface swaps, however, the numbers do no change drastically. 

As previously mentioned, there are two costs measured here - the price impact on the pool and gas costs. The grey line in the figure above charts the breakeven swap size if gas was always always the median cost over the period. This shows us fluctuations due to price impact differences, removing the impact of gas. We can see that while the differences in price impact from Ethereum to Arbitrum are a major driver of the breakeven trade size, but we can see other fluctuations due to differences in gas costs.

If drawbacks for L2s are addressed appropriately, then there may be no reason for Ethereum to have significantly more TVL than an L2. With gas costs lowering by an estimated 20x-60x the current rate on rollups due to EIP-4844 introduced in Dencun hardfork, this may further increase the breakeven swap amount. If liquidity differences between L2s and mainnet converge, there may not be any trade that is better on Ethereum mainnet.\footnote{\href{https://blog.oplabs.co/eip-4844-an-optimistic-bet-on-rollup-scalability/}{Source for 20x} and \href{https://twitter.com/VitalikButerin/status/1764139239233749047}{source for 60x}}

\medskip
\paraheader{LPs Increased Concentration leads to greater capital efficiency.} 

We have shown that smaller swappers get better execution on Uniswap v3 due to lower fixed gas costs from L2s. However, perceived security of the chain and subpar UX limit the amount of capital that is currently deployed on these chains. 

One of the major critiques of Uniswap v2 was that it created capital inefficiencies by locking capital at prices it was unlikely to ever trade at. Uniswap v3 changed this by allowing the concentration of liquidity around desired price intervals. However, a more concentrated position is more likely to require rebalancing as the price of the pool moves. Rebalancing incurs a gas cost, meaning that when the gas costs are higher, efficient users will rebalance less. On mainnet, this means capital is still not as concentrated as it could be if gas had no cost. We could see more concentration on L2s, where lower gas prices mean the fixed costs for LPing are lower.

The data supports the theory that as LPs are able to adjust their placement of liquidity more cheaply on L2s, there is more concentration around the midmarket price. Because of gas costs, L2s have at least 75\% more concentration than Ethereum. There are also significantly more liquidity positions opened on L2s, showing that users are rebalancing their positions more frequently. Not only that, we see that there are more unique liquidity providers on L2s, meaning users may be priced out of providing liquidity on Ethereum.

\begin{figure}[H]
\centering
\textbf{Figure 3: Median liquidity concentration around mid-market of ETH/USDC
}\par\medskip 
\includegraphics[width=.8\textwidth]{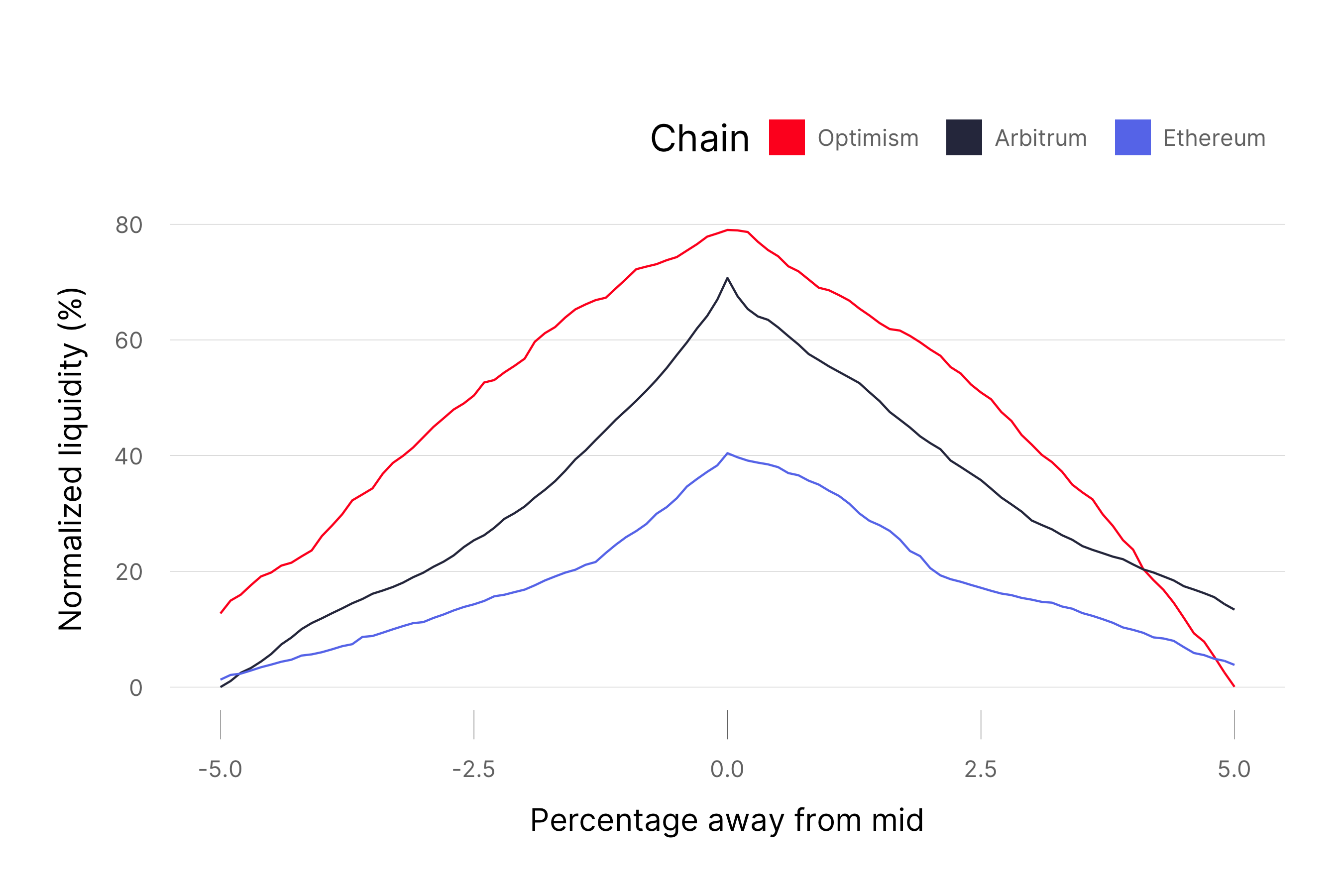}
\caption*{Note: We calculate the liquidity distribution around the current mid-price every 15 minutes from 03-01-2023 to 01-01-2024 and take the normalized liquidity distribution around mid for the studied chains.}
\end{figure}

Normalizing liquidity at each sample period, we calculate that median liquidity at the current mid market price is 70\% of the total on Arbitrum and 80\% of total on Optimism. On Ethereum, it is only 40\%. This means that in the median sample, L2s have at least 75\% more concentration at the pool midpoint than Ethereum. 

\begin{figure}[H]
\centering
\textbf{Table 2: Uniswap v3 LP Positions Selected Statistics
}\par\medskip 
\includegraphics[width=.8\textwidth]{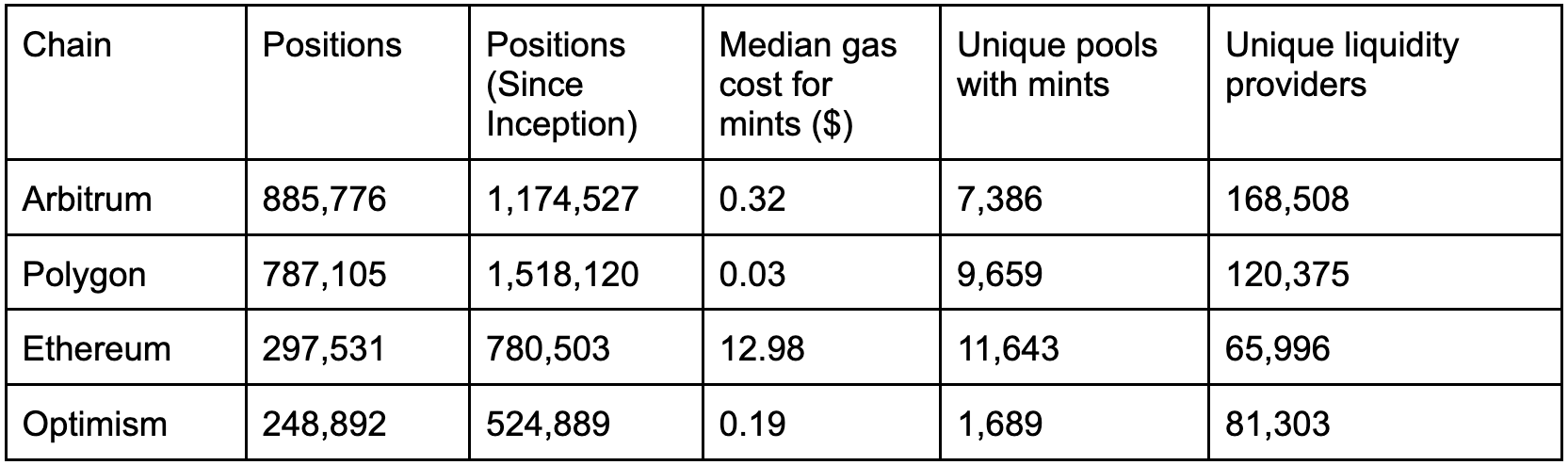}
\caption*{Note: Everything calculated as of end of 2023 except where marked.\\
\href{https://dune.com/queries/3290039}{Source}}
\end{figure}

There is also evidence that users are rebalancing more frequently, resulting in more efficiently used concentrated liquidity. More concentration of liquidity makes capital more efficiently used on the Uniswap Protocol, as less capital is locked up far away from the current midpoint. When a position is rebalanced in Uniswap v3, technically a new position is created, meaning that more frequent rebalancing creates more unique positions. In \textbf{Table 2}, we see significantly more positions created on L2s. Arbitrum has almost three times more positions created this year than Ethereum - a similar trend in Polygon as well.  We see that users (likely sophisticated) are rebalancing more frequently. 

However, we find that it is not just that the same set of users are providing more concentrated liquidity. In \textbf{Table 2}, we see that more unique wallets are provide liquidity, as there are almost three times the unique wallets providing liquidity on Arbitrum. This could indicate that many would-be LPs are priced out by the expensive gas cost of liquidity provision on mainnet Ethereum. 

To zoom in on a specific pool, we will focus on arguably the most important pool on Uniswap v3, the 5 bps ETH/USDC pool. Since, most onchain pools are paired tokens paired against (W)ETH, this is the pool used to utilize stablecoins in Defi via multi-hop trades through WETH/USDC. 

\begin{figure}[H]
\centering
\textbf{Table 3: ETH/USDC 5 bps Selected Statistics YTD
}\par\medskip 
\includegraphics[width=.8\textwidth]{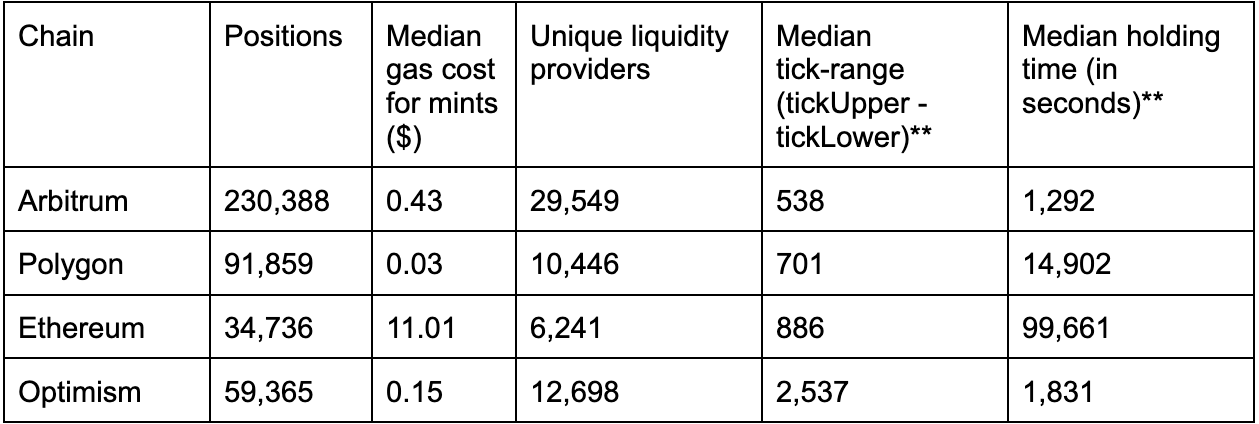}
\caption*{Note: ** excludes JIT LPs/Limit Orders (tick-range = 1 tick spacing) \\
\href{https://dune.com/queries/3294748}{Source}}
\end{figure}

While all the 5 bps ETH/USDC pools have similar dynamics, we will focus on Ethereum vs. Arbitrum. One notable difference between Ethereum and Arbitrum is the average holding time of each position. Ethereum holding time is more than 75x longer from first mint to first burn than Arbitrum, meaning that positions are rebalanced less frequently. There are fewer liquidity providers on Ethereum mainnet, they are holding their positions longer, and their positions are wider.

As a result, there are ultimately fewer but larger positions on Ethereum. This could be due to perceived security risks inherent to L2s, as even a low chance of total loss of funds - due to a security issue - must be internalized when making trading decisions. This low chance causes large LPs to utilize less capital on L2s, creating smaller liquidity positions on average. We will discuss these tradeoffs more in later sections. 

\medskip
\paraheader{Increased Fee Returns For Liquidity Providers.} 

We have already shown the positive effects on market structure for LPs from lower gas - increased LP concentration, more unique liquidity providers, and decreased holding times. A second important aspect of the studied chains are their shorter block times. Historically, it has been difficult to calculate the benefit for market structure changes on accured LP fees, as LPs, arbitraguers, and swappers may respond differently to market structure changes. LPs are especially difficult due to utilizing different portfolio benchmarks. Liquidity providers may have arbitrary reasons for providing liquidity, and may benchmark against holding an arbitrary portfolio of tokens (impermanent loss), loss vs rebalancing (\LVR), or any other arbitrary strategies. However, we will focus on the well-researched \LVR, established in \cite{milionis2023automated}, to compare between L1 and L2 LP performance. We also choose this because improving \LVR generally improves returns on other strategies, and because the benchmark for the portfolio is the same between chains and at arbitrary times.

In the \LVR with fees framework, arbitrage losses increase with the square root of block time. This is because the LP fee of the pool creates a lower bound for price movements in which arbitrage transactions are not profitable, described as the no-trade region. Shorter block times means there is less time for the market price of the assets to move at a given level of volatility and therefore a higher likelihood that arbitrage transactions on that price movement will not be profitable (due to staying within the no-trade region).

When using Fees - \LVR to compare full-range positions (or any identical LP strategy), the costs of hedging the LP position are the exactly the same. This is because the path of the asset and the hedge occurs outside of the decentralized exchange by assumption , meaning for like assets and like LP strategies, the LVR cost is identical. Because the daily LVR cost is identical between all the studied pools, we are able to simplify the comparison to just the fee returns of the LPs, which can be found directly from swap transactions. This allows a simple comparison of Fees - \LVR utilizing only change in fees.

From \cite{mitchell2022layer2}, we can see there is very little difference between \textbf{aggregate} ETH/USDC fee returns of full-range positions on L1s and L2s. We know that there is more retail flow transacting on Ethereum, which may offset the decreased fee returns from arbitrage. We have also shown that retail traders continually trade on Ethereum mainnet - even though this is not optimal, likely because of lagged adoption of other chains. However, because of the coordination required between LPs and swappers, liquidity providers may stay on Ethereum because the fees from retail flow offset the increased costs from arbitrage (even if pareto optimal for both parties to be on L2s). Retail traders are the bulk of positive returns for liquidity providers, so liquidity providers must stay where they are. 

Additionally, retail trading fees are hard to model and their reasons for trading are not well understood. We believe as adoption of cheaper chains increased, users will either move to cheaper but decentralized chains, or the choice will be abstracted away through market structure innovations like Crosschain UniswapX. Accordingly, we want to present data that attempts to quantify the impact of changing block times on arbitrage fees from LPs. To focus on the impact of arbitrage of lower costs and shorter block times, we will remove retail transactions, as arbitrage fee returns react instantly to market structure changes.

To segment arbitrage fee returns, we utilize router addresses for all retail flow on Arbitrum, Ethereum, and Optimism. Since Uniswap v3 requires a contract to interact with it (and cannot be called directly by an end user's wallet) and contract deployments cost several magnitudes more than a swap, it feasible to check these contracts by hand. For example, on Arbitrum, we checked around 300 contracts (68 of which are interfaces) spanning more than 95\% of volume and transactions. For context, around 1 million wallets have swapped on Arbitrum. The transparent nature of the blockchain also makes it feasible to see what these contracts are doing to ensure they are being used as expected. While we likely missed a very small number of interfaces, we believe this sample well-spans the collection of retail interfaces.\footnote{If anything, we are biasing against L2s in this methodology, as gas optimizations are less meaningful on L2s, meaning some searchers utilize Uniswap Interface contracts for arbitrage on L2s.} The transaction counts and volume also follow a power law distribution when segmented by originating contract, and we went from largest first. For replication, a list of all the tracked contracts can be found in \href{https://github.com/aadams/layer2-be}{this repository}.

Next, we remove all swap transactions that originate with one of these interface routers (to remove retail flow). Lastly, we take all the remaining arbitrage transactions left, re-calculate their impact on state, and generate the full-range fee returns from all of these swaps. This is utilizes the feeGrowthGlobal values found in Uniswap v3 state. For proofs and instructions on replicating this methodology, see \cite{adams2022fees}.

\begin{figure}[H]
\centering
\textbf{Figure 4A: Distribution of daily fee returns by token on Arbitrum}
\par\medskip 
\includegraphics[width=.9\textwidth]{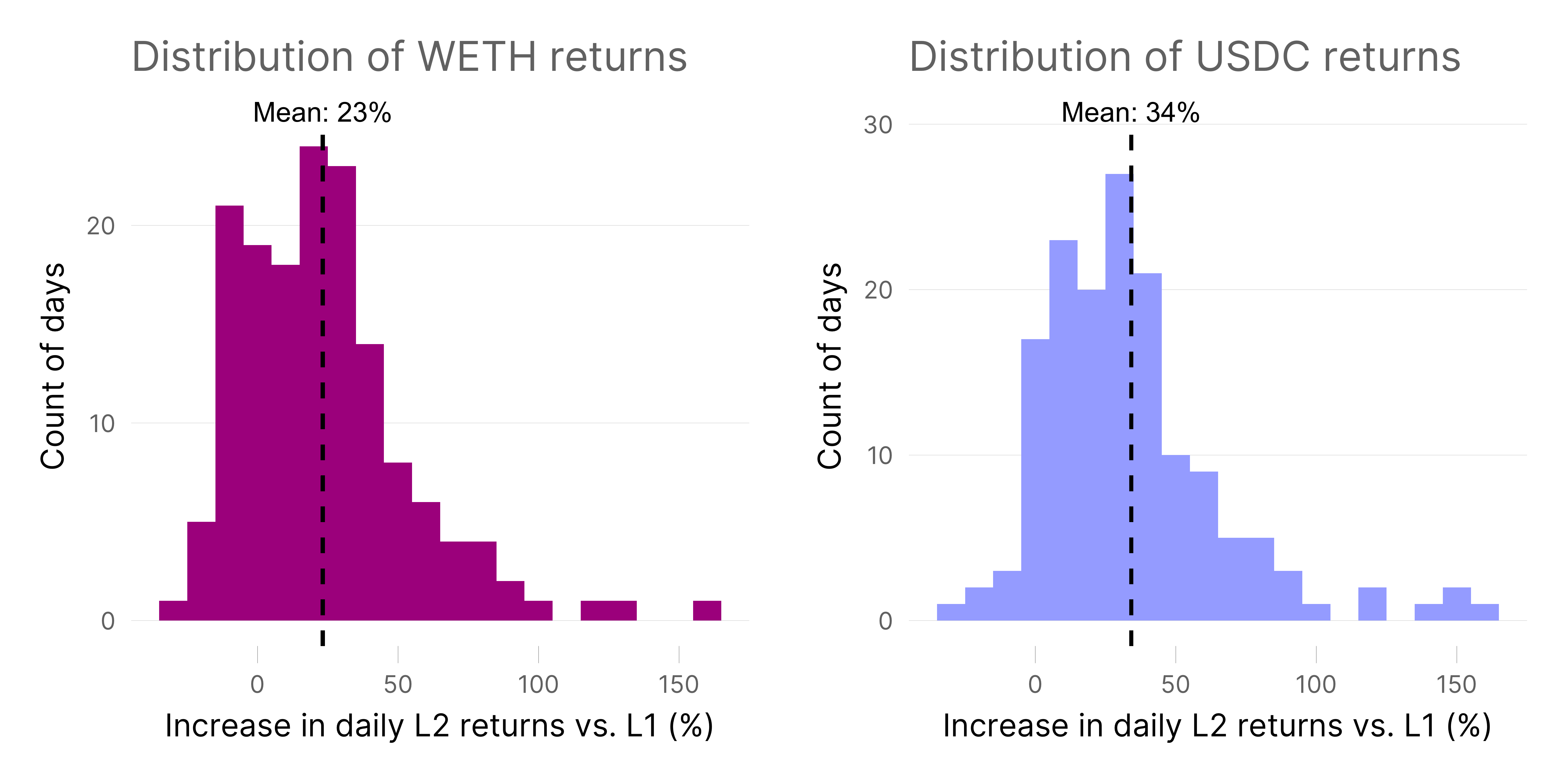}
\end{figure}

\begin{figure}[H]
\centering
\textbf{Figure 4B: Distribution of daily fee returns by token on Optimism
}\par\medskip 
\includegraphics[width=.9\textwidth]{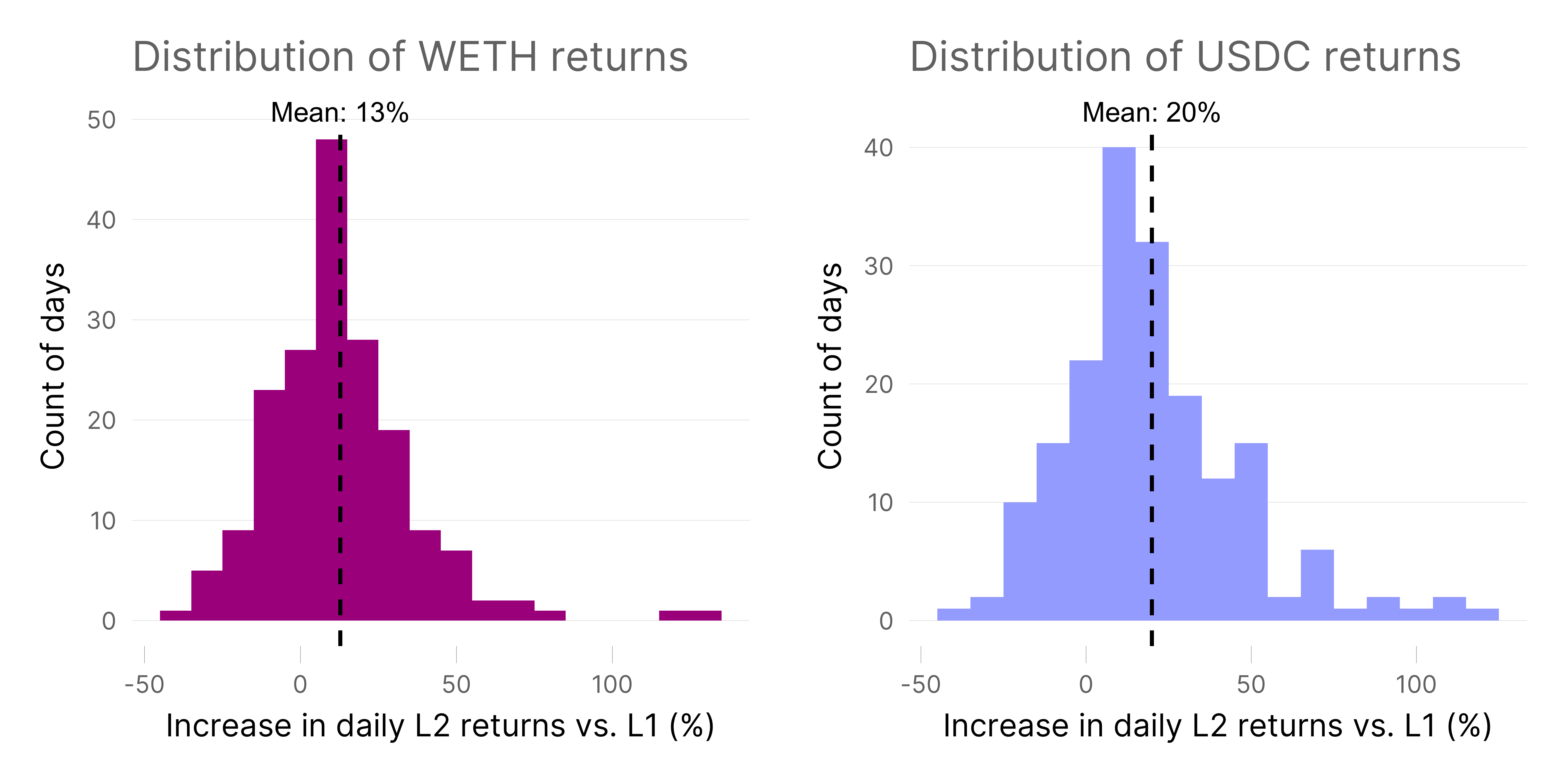}
\caption*{Note: Data calculated from March 1st, 2023 to December 1st, 2023}
\end{figure}

Calculations show that full-range LPs on L2s make an average of 20\% more returns from arbitrage compared to the same day on Ethereum mainnet. 

As previously stated, the increase in fee-returns is from gas cost differences and shorter block times. However, during this sample period (done purposefully), Optimism and Arbitrum had nearly exact gas costs, while block times were the main difference. To zoom in on like-samples, we focused on March 1st, 2023 to December 1st, 2023. Over this period, we find that both Arbitrum and Optimism outperformed Ethereum by a significant margin - around 20\% on average. Finally, by focusing only on arbitrage transactions (which will execute at any depth if profitable), we can remove the impact of liquidity differences of these pools between chains, meaning that equalizing the liquidity on L1s and L2s will not impact these dynamics. There is an interaction between full-range fee returns and increased concentration of other LPs, but it is minimal and due to gas costs. A deep dive into intra-pool LP dynamics is out of the scope of this paper - for that, see \cite{milionis2023flair}. 

Since September 2021, Ethereum mainnet returns for ETH/USDC would increase by about \$20m, surpassing some calculated losses to LPs during that time - meaning that Ethereum LPs on ETH/USDC could be in profit from only this change if they mark their returns with 5 minute \LVR.\footnote{See here for \href{https://dune.com/queries/3380200}{query} and here for \href{https://www.gauntlet.xyz/resources/uniswap-user-cohort-analysis}{data sources} and \cite{canidio2023arbitrageurs}{here}} This assumes that Ethereum adopts the same block production mechanisms as L2s today (or that all capital and flow is perfectly migrated to L2s). For example, the FM-AMM from \cite{canidio2023arbitrageurs} and associated CowSwap implementation on Ethereum mainnet \href{https://x.com/CoWSwap/status/1757783798601986400?s=20}{claim increasing fees} by 5-7\% fixes \LVR, while simply using faster and cheaper chains increases it by an estimated 10-15\%.

Importantly, the numbers for Ethereum include fee returns from sandwich transactions - which accounts for around 10\% of ETH/USDC returns according to \cite{holloway2023value} - meaning the difference is probably larger than 20\%. Because sandwich bots utilize contracts which are generally also arbitrage bots, our methodology will include them as arbitrage transactions. Because of the current nature of sequencers, there are fewer sandwiches to no sandwiches on L2s, as sandwich trades without strict control of block ordering run the risk of being unbundled.\footnote{For MEV number, see \href{https://dune.com/hildobby/sandwiches?Blockchain_e8f77a=arbitrum}{here} and for a description on \href{https://ethresear.ch/t/equivocation-attacks-in-mev-boost-and-epbs/15338}{unbundling}} Accordingly, this removes sandwich transactions on the studied L2s, but also their associated LP returns. 

In our sample period, Arbitrum had larger arbitrage fee returns than Optimism while gas fees were similar. For LPs, the practical difference between the chains is block times. As said previously, Arbitrum built blocks on demand, allowing for multiple arbitrage transactions to happen in a 2 second window - possibly hinting that \textbf{2 seconds is still too large of a block time for optimal level of fee returns}. Finding the optimal block time is likely dependent on volatility and is not possible in our current model, thus it is out of the scope of this paper.

\section{Drawbacks for Uniswap v3}

We have talked extensively about the benefits of faster and cheaper for users, but current L2s also have drawbacks.

First, the sequencing of transactions is currently chosen by privileged actors, meaning that sequencers for these rollups are currently centralized.\footnote{We are using the term rollup to denote that Polygon does not have this issue} Currently, rollups are designing ways to decentralize block ordering to a distributed committee of sequencers, but these solutions are not in production at the moment. By having one centralized sequencer, sequencers could take advantage of their status in the system to generate MEV profits by re-ordering transactions - similar to how PoW miners or PoS validators extracted value. However, in practice, sequencers to date haven’t done this and prefer known/fair block ordering methods like priority gas auctions or first come, first served. As previously stated, the current nature of sequencers means less sandwiches from non-privileged parties, due to the risk of being unbundled and shielded mempools. The MEV on L2s is largely currently confined to non-adversarial MEV (arbitrage, back-running, liquidation) due to this. MEV extraction has also led to centralization and instability in block production, which might not preferable for L2s.\footnote{To see issues with MEV extraction, see \href{https://ethresear.ch/t/execution-tickets/17944}{here} and \href{https://ethresear.ch/t/timing-games-implications-and-possible-mitigations/17612}{here}}

Second, the distinct ecosystems of L2s create liquidity fragmentation. In their current design, L2s cannot trustlessly speak to each other natively in real-time. This means that liquidity on different L2s are not composable. Put another way,  you cannot natively trade on one L2 and use the liquidity immediately on the other. Currently, each chain must bootstrap distinct liquidity sources for their own rollup, which duplicates capital locked up to facilitate trading, adds additional costs for LPs due to arbitrage, and increases price impact for users. However, there are a few mechanisms which could facilitate interaction between L2s, such as bridging systems like Wormhole or intent-based trading systems like Cross-chain UniswapX.\footnote{For a description of UniswapX, see \href{https://uniswap.org/whitepaper-uniswapx.pdf}{here}}

Third, some of the current optimistic rollups lack fraud proofs, which are infrastructure to trustlessly recover from sequencer errors or incorrect transactions.\footnote{This is not broadly true. Arbitrum has working fraud proofs for whitelisted actors} Because optimistic rollups assume by default that transactions are valid, sophisticated actors need to utilize fraud proofs to show that posted transactions/blocks are invalid. After seven days, current state is assumed to be correct - meaning transactions like withdrawals are processed. If an invalid withdrawal is created and not disputed within seven days, then invalid transactions will be canonical and the withdrawal is processed. On Optimism, fraud proofs allow the trustless and retroactive dispute of a transaction from the canonical state roots, effectively removing it from the chain trustlessly if the fraud proof is valid. On Arbitrum, if there are conflicting state root proposals, validators stake on histories - effectively utilizing the the fraud proofs to solve the fork choice. For rollups, fraud proofs are a powerful unlock for decentralization of the chain as a whole but have implementation challenges. However, significant work is going into productionalizing fraud proofs.

Lastly, bridging between ecosystems can be costly and time consuming. To use the native bridges between chains, users must return to mainnet and then enter the canonical bridge for the other ecosystem. On optimistic rollups, it takes seven days to bridge funds back to mainnet, and depositing into a rollup costs around 8\$ at average gas prices. This adds both time and costs to transactions that are generally time sensitive. As previously said, there are methods to circumvent this like Cross-chain UniswapX. 

There is evidence that users internalize these implicit costs when interacting with L2s. \cite{chemaya2022cost} estimated that L2 users internalize a 3.29\% probability of total loss when trading CPMMs on Optimism and 0.68\% on Polygon. They utilize the actual revealed preference of users switching between mainnet and the L2s to calculate this value. However, over time this number is trending towards 0, as the security tradeoffs of these systems lessen through the aforementioned improvements. The authors do note that other factors could impact this parameter, such as community, user experience, and overall integrations. We find similar results, as smaller users continue to execute trades on Ethereum mainnet despite the relatively better execution on L2s.

On the other hand, \cite{cong2023scaling} finds L2s may lead to more secure markets, because fixed costs for interacting with markets are lower. Looking at the accuracy of Chainlink oracle feeds, Cong et al. estimate that costs for operators dropped by 76\% while each round had 79\% more operators, leading to a more secure, yet cheaper, feed. We find similar results with more unique liquidity providers providing liquidity on Uniswap v3 and smaller average Uniswap Interface trade size  - signaling that users may be priced out of interacting with the mainnet ecosystems due to high gas costs.

\section{Conclusion}

Overall, we find that faster and cheaper chains positively impact market structure on decentralized markets compared to Ethereum mainnet. In sum, swappers benefit from lower gas cost, more concentrated liquidity, and prices that update more frequently. LPs benefit from cheaper position update costs in concentrated liquidity, more arbitrage flows resulting in higher profit, and lower cost of pool deployment. 

As L2 costs continue to plummet for all users, the benefits of decentralized markets will continue to shine, while removing negative barriers. The design of current decentralized markets may also markedly change, as many of the jarring constraints for these markets are due to high costs for end users. Increasing a trade's cost by double when the underlying cost is 1c is proportionally less economically impactful than doubling the cost of a \$10 translation. For decentralized markets to fulfill their full potential, aggregate trading costs must continue to decline and user experience must continue to improve.

We expect that users will continue their crosschain migration as perceived and realized security of rollups continues to increase, decreasing the main benefits of continued swapping on Ethereum mainnet. This is in the best interest of both Ethereum and L2s, as this will dramatically decrease the cost and increase throughput of the entire Ethereum network.

In this paper, we only studied a subset of the growing number of chains. Because we mainly studied the most active chains today, we make no attempt to show trading benefits of newer technologies like ZK-proofs, other alternative chains, and the future benefits for the studied L2s due to smaller Uniswap v3 integrations.

Uniswap v3 was also optimized for Ethereum mainnet gas costs, meaning that the trading experience for the forthcoming Uniswap v4, described in \cite{Adams23}, may benefit greatly from cheaper chains. Uniswap v4 allows for customization on top of the Protocol layer, which is up to the developer. Integrators on cheaper chains can utilize more complicated hooks, because of lower gas costs, to improve user experience. We have seen this already with hooks utilizing \href{https://github.com/saucepoint/v4-axiom-rebalancing/}{zero-knowledge proofs} and \href{https://uniswaphooks.com/components/hooks/unisuave-ethglobal}{Flashbots’ SUAVE}, which were deployed on L2s and are likely not cost effective on mainnet.

However, we believe that the studied generalized L2s still have many benefits that users can utilize today, and any future improvements will only continue to benefit the trading experience.

\newpage

{\small
  \singlespacing
  \bibliographystyle{ACM-Reference-Format}
  \bibliography{references}
}

\end{document}